\documentclass[twocolumn,showpacs,preprintnumbers,amsmath,amssymb]{revtex4}

\usepackage{graphicx}
\usepackage{dcolumn}
\usepackage{bm}

\begin{document}

\preprint{}

\title{Complexification of The Vacuum and The Electroweak Gauge Symmetry }

\author{Alp Deniz \"Ozer}
\email{oezer@theorie.physik.uni-muenchen.de}
\affiliation{%
Ludwig Maximilians University  Physics Section, Theresienstr. 37,
80333  Munich Germany \\
}%

\date{\today}

\begin{abstract}
In this paper we investigate the consequences of imposing an
$SO(4)$ symmetry both on the scalar field and on spacetime prior
to any spontaneous symmetry breakdown in the electroweak model.
First we argue that the local electroweak gauge symmetry is
implied by the initial $SO(4)$ symmetry of the scalar field. Then
we parameterize the vacuum, in a slightly different way with
respect to  the conventional assignment in the electroweak model,
around the ground state mainly in two different ways: Once over
the local electroweak gauge transformations and once over the
local $\gamma$ basis which generates the $SO(4)$ symmetry of the
scalar field. Through comparing the two, we conclude that a
consistent parametrization, requires the complexification of the
scalar field, which simultaneously requires spacetime to obey
local Lorentz invariance in the broken phase of the electroweak
vacuum. After spontaneous symmetry breakdown and complexification,
the $SO(3,1)$ invariant quantity of the scalar field is identified
to be the Higgs mass, which yields $m_H=v/2 \cong 123$ GeV. Also
the vacuum state is normalized to a unit state by rescaling the
local $\gamma$ basis, so that the metric tensor inherently
contains the speed of light in the Higgs vacuum. We shortly
consider the validity of the $SO(3,1)$ invariance property of the
vacuum state in terms of the Goldstone modes, in case the Higgs
field departs from the ground state. The main goal of the paper is
to treat spacetime and vacuum as the two faces of the same
medallion, and to relate the local spacetime symmetry with the
electroweak gauge symmetry, both in the broken and unbroken phases
of the vacuum via complexification and spontaneous symmetry
breaking.
\end{abstract}

\pacs{11.30.Cp, 11.15.Ex, 14.80.Bn}
\keywords{Complexification,Vacuum, Higgs mass} 

\maketitle

\section{Introduction}
The current formulation of the electroweak gauge theory inevitably
requires the existence of a scalar field for a few fundamentally
important reasons. First of all, the electroweak gauge
interactions among elementary particles naturally occur in vacuum,
which is understood to be the {\it stable} ground state of the
scalar field. Historically, to our knowledge, the scalar field has
not been incorporated in the electroweak gauge
theory\cite{Weinberg:1967tq}\cite{Salam:1968rm}\cite{Glashow:1970gm}
to provide an underlying physical continuum for gauge
interactions. The main motivation was to utilize the {\it physics}
in the spontaneous breakdown of the scalar
field\cite{Goldstone:1961eq}\cite{Goldstone:1962es}, since it
provided masses for the mediators in a gauge invariant and
renormalizable way\cite{'tHooft:1972fi}, and also explained the
non conservation of isotopic spin.

From the other side, a general fact in nature is that interactions
among matter and fields are solely governed by gauge theories,
which come along with their respective gauge symmetries\cite{Fritzsch:2002jv}.
We also demand gauge theories to be locally Lorentz invariant
since initial and final particle states are expressed through
quantities like energy and momentum\cite{Eidelman:2004wy} which
obey the laws of special relativity. We know that Lorentz
Invariance is tightly connected with the abstract notion {\it
space-time}, which makes hardly physical sense unless we consider
matter and fields attached to it. The remarkable thing about
spacetime, matter and fields is that it constitutes in a
completely different fashion the ingredients of special and
general relativity. In contrast to the electroweak theory,
gravitational interactions or in other words the motion of massive
particles governed by the laws of general relativity are
accommodated by {\it spacetime}.

If it is correct that the Higgs
mechanism\cite{Higgs:1964ia}\cite{Higgs:1964pj} is the ultimate
formalism to create inertial and gravitational masses for
elementary particles, we should somehow be able to relate the
scalar field, which is responsible for the mass generation
mechanism, with {\it spacetime}. In this paper we will treat
vacuum and spacetime as the two faces of the same medallion and
will draw certain consequences out of it, such as the Higgs boson
mass and speed of light in the Higgs vacuum. The relation between
the two faces will be step wise introduced through out the work,
the main steps however are summarized in the conclusion part for
clarity. As a starting point it would be most appropriate to
consider the global transformation properties of the 4 component
scalar field.

Let us start by placing four {\it real} valued scalar fields in a
column
\begin{equation}\label{vector}
\phi = \frac{1}{\sqrt{4}}\left( \begin{array}{c}
  \phi_1 \\
  \phi_2 \\
  \phi_3 \\
  \phi_0 \\
  \end{array} \right)
\end{equation}
A Lagrangian of the scalar field, in the four vector
representation that exhibits a {\it global} $SO(4)$ symmetry, is
\begin{equation}\label{scalarfield}
 L_\phi=(\partial_{\mu} \phi)^T (\partial^{\mu} \phi)+\mu^2 \left(\phi^T \phi\right)-\lambda \left( \phi^T
 \phi\right)^2
\end{equation}
where $\mu^2 , \lambda$ are initially chosen to be positive.  We
characterize the vacuum through these scalar fields. The vacuum
state corresponds to the ground state of the scalar field. Prior
to any spontaneous symmetry breakdown and phase transition the
null vector $\phi \equiv 0$, obviously defines a {\it false}
vacuum since it pertains to an {\it unstable} ground state. This
false vacuum is trivially $SO(4)$ invariant. Let us {\it
additionally} demand that our $4$ dimensional
spacetime\footnote{Instead of 4 dimensional spacetime, it would be
appropriate to call it 4 dimensional space only, so that time is
treated on equal footing with other coordinates.} exhibits a
Euclidean signature and is subject to the same symmetry of the
scalar field.

The real generators $\Sigma_{ab}$, that satisfy the $SO(4)$ Lie
Algebra, can be generated through
\begin{equation}
  \left( \Sigma_{ab} \right)_{ij}  = - \left( \delta_{ai} \, \delta_{bj} \, -\delta_{ji} \,
  \delta_{ab}\right)
\end{equation}
\begin{equation}
  \left[  \Sigma_{ab}, \Sigma_{cd}  \right]  = \Sigma_{bc} \, \delta_{ad}+\Sigma_{ad}
  \, \delta_{bc}-\Sigma_{ac} \, \delta_{bd}-\Sigma_{bd} \,
  \delta_{ac}
\end{equation}
The above real valued Lie Algebra of $SO(4)$,  can be made complex
by allowing $\Sigma \rightarrow i \, \Sigma$ as below. But in
general, this complex valued Lie Algebra is satisfied by the so
called spinorial representation of $SO(4)$. The generators of this
spinorial representation can be produced by a set of $4$ gamma
matrices :
\begin{equation}
  \left[\Sigma_{ij}, \Sigma_{kl}\right]=  -i\, \left(\Sigma_{jk} \, \delta_{il}+\Sigma_{il}
  \, \delta_{jk}-\Sigma_{ik} \, \delta_{jl}-\Sigma_{jl} \, \delta_{ik}\right)
\end{equation}
\begin{equation}\label{so4}
 \Sigma_{ij} = \frac{i}{2} \left[ \gamma_i, \gamma_j \right] \ , \ \ \ \
2\, \delta_{ij} = - \, \{\gamma_i, \gamma_j \}
\end{equation}
where $\delta_{ij}$ carries a Euclidean signature for $i,j,k,l $
$=\{0,1,2,3\}$ such that $\delta_{kk}=1,$ and the curly brackets
denote anticommutation. Let us consider as next the gamma matrices
of the chiral representation. These are explicitly stated as
\begin{equation}
 \gamma_{i}= \left(
 \begin{array}{cc} 0 & \sigma_{i} \\
 -\sigma_{i} & 0  \end{array}
 \right)  \ \ , \ \ \
   \gamma_{0}= \left( \begin{array}{cc} 0 &  -I \\
  -I & 0  \end{array} \right)
\end{equation}
\begin{equation*}
 \gamma_{5}=
 \left( \begin{array}{cc} I & 0 \\
 0 & -I  \end{array} \right)
\end{equation*}
where $i=1,2,3$  and $\sigma_i$ are the Pauli spin matrices. The
chiral representation composed of $\{\gamma_0 , \gamma_1,
\gamma_2, \gamma_3 \}$ satisfies  Lorentz covariance hence belongs
to  $SO(3,1)$ and has the  following signature
\begin{equation}\label{set1}
\begin{split}
    (\gamma_0)^2=1 \, \ \ \ \   \ \ \  (\gamma_1)^2=(\gamma_2)^2=(\gamma_3)^2=-1 \\
\end{split}
\end{equation}
If we replace $\gamma_0$ by $-i\gamma_0
\stackrel{\text{def}}{=}\tilde{\gamma}_0$. The signature turns out
to be Euclidean
\begin{equation}\label{set2}
\begin{split}
    (\tilde{\gamma}_0)^2=-1 \, \ \ \ \    (\gamma_1)^2=(\gamma_2)^2=(\gamma_3)^2=-1 \\
\end{split}
\end{equation}
Now we can use the basis $\{\tilde{\gamma}_0 , \gamma_1, \gamma_2,
\gamma_3 \}$ in eq. (\ref{so4}), so that the generators produced
by this basis satisfy the $SO(4)$ Lie Algebra.  Let us construct a
sum over the scalar fields spanned by the basis
$\{\tilde{\gamma}_0 , \gamma_1 , \gamma_2, \gamma_3 \}$. We obtain
\begin{equation}\label{thefour}
\phi=\frac{\phi_i \gamma_i}{4}+ \frac{\phi_0 \tilde{\gamma}_0}{4}=
  \frac{1}{\sqrt{8}}\left[ \begin{array}{cccc}
  0 & 0 & \phi^0_1 & \phi^{+}_1  \\
  0  & 0 & \phi^{-}_2 & -\phi^0_2 \\
 -\phi^{0}_2  & -\phi^{+}_2 &  0 & 0  \\
 -\phi^{-}_1 & \phi^{0}_1 & 0 & 0 \\
 \end{array} \right]
\end{equation}
\begin{equation}
\phi^c = -\frac{\phi_i \gamma^*_i}{4}-\frac{\phi_0
\tilde{\gamma}^*_0}{4}=
  \frac{1}{\sqrt{8}}\left[ \begin{array}{cccc}
  0 & 0 & -\phi^0_2 & -\phi^{-}_2  \\
  0  & 0 & -\phi^{+}_1 & \phi^0_1 \\
 \phi^{0}_1  & \phi^{-}_2 &  0 & 0  \\
 \phi^{+}_1 & -\phi^{0}_2 & 0 & 0 \\
 \end{array} \right]
\end{equation}
Here the upper indices $\{+,0,-\}$ are showing the charges with
respect to the diagonal generator $\Sigma_{12}$ which is obtained
from
\begin{equation}
\Sigma_{ij} = \frac{i}{2} \left[ \gamma_i, \gamma_j \right] =
\epsilon_{ijk}\left(
\begin{array}{cc}
  \sigma_k & 0 \\
  0 & \sigma_k \\
\end{array}
\right)
\end{equation}
The lower indices in $\{\phi^0_1,\phi^0_2 ,\phi^{+}_1,\phi^{-}_2
\}$ are introduced just to distinguish among the entries. The
explicit expressions of these 4 fields in terms of
$\{\phi_1,\phi_2 ,\phi_3,\phi_0 \}$ is found to be
\begin{equation}\label{complexfields}
\begin{array}{cc}
  \phi^0_1    & =  (\phi_3 - i \  \phi_0 )/\sqrt{2}  \\
  \phi^0_2    & =  (\phi_3  + i \ \phi_0 )/\sqrt{2}  \\
\\
  \phi^{+}_1    & =  (\phi_1 - i \ \phi_2)/\sqrt{2}  \\ 
  \phi^{-}_2    & =  (\phi_1 + i \ \phi_2)/\sqrt{2}  \\ 
\end{array}
\end{equation}
From the above expressions it is easy to see that the imaginary
sign in front of  $\phi_0$ in both  $\phi^0_1$ and $\phi^0_2$
stems from  $\tilde{\gamma}_0$ $=$ $-i\gamma_0$. If we had not
utilized $\tilde{\gamma}_0$ in the expansion, the fields
$\phi^0_1$ and $\phi^0_2$ wouldn't have had a {\it complex form}.
An explicit imaginary sign was required to appear in front of
$\phi_0$, in this respect.
Also note that $Tr[\phi \phi]$ is thereby an $SO(4)$ invariant.\\
\begin{equation}\label{inv}
  -Tr[\phi \, \phi] = \frac{1}{4}\left(
  \phi^2_1+\phi^2_2+\phi^2_3+\phi^2_0\right)
\end{equation}
The Lagrangian of the scalar field can equivalently be reexpressed
through $\phi$ in eq.(\ref{thefour}) as
\begin{equation}\label{scalar-Lag}
  L_\phi =  -Tr\left[\, \partial_{\mu} \phi \,  \partial^{\mu} \phi \, \right]
  -\mu^2 Tr\left[\, \phi \, \phi\, \right] - \lambda  \, Tr \left[ \, \phi
 \, \phi \, \right]^2
\end{equation}
There is no Hermitian conjugation in the product $\phi \, \phi$.
Note that the $\gamma$  matrices can also  be treated
algebraically, so that the trace operation can be dropped as well.

In Eq.(\ref{complexfields}) there are only four independent
fields. These fields can be  organized into two doublets.
\begin{equation}\label{doublet+0}
\phi = \left( \begin{array}{c}
  \phi^+_1 \\
  \phi^0_1 \\
\end{array} \right) =
\frac{1}{\sqrt{2}}\left( \begin{array}{c}
  \phi_1 - i \phi_2\\
  \phi_3 - i \phi_0\\
\end{array} \right)
\end{equation}
and
\begin{equation}\label{doublet-0}
\phi^* = \left( \begin{array}{c}
  \phi^-_2 \\
  \phi^0_2 \\
\end{array} \right) =
\frac{1}{\sqrt{2}}\left( \begin{array}{c}
  \phi_1 + i \phi_2\\
  \phi_3 + i \phi_0\\
\end{array} \right)
\end{equation}
The above doublets are complex conjugates of each other, since
$\{\phi_1,\phi_2 ,\phi_3,\phi_0\}$ are real fields. Note that they
transform under $\Sigma_{ij}$, for $i,j=\{1,2,3\}$ independently,
so that we can consider them separate $SU(2)$ doublets. The nice
thing about the expansion in Eq.(\ref{thefour}) is that it allows
each doublet to be  grouped so as to include a copy of all four
scalar fields. We see that the $\{+,0\}$ charges in $\phi$, do not
correspond to the $\sigma_3$ charges of $SU(2)$ but are due to
$\Sigma_{12}$. We will come back to this point again.

From the other side the Lagrangian of the scalar field in
eq.(\ref{scalarfield}) can sufficiently be reproduced  by any of
the above doublets, which are indeed equivalent. Since we have put
the four fields into a complex representation, the Lagrangian
should then be written in a complex form
\begin{equation}\label{scalarfield-c}
  L_\phi=\frac{1}{2}(\partial_{\mu} \phi)^\dagger (\partial^{\mu} \phi)+
  \frac{\mu^2}{2} \left( \phi^\dagger \phi\right) -\frac{\lambda}{4} \left( \phi^\dagger
 \phi\right)^2
\end{equation}
Such a restatement imposes an overall $U(2)$ symmetry on the
Lagrangian, which is nothing but an $SU(2) \times U(1)$ symmetry.
The above $\{+,0\}$ charges in $\phi$ can be retrieved from this
$SU(2) \times U(1)$ symmetry, such that
\begin{equation}
    \frac{\sigma_3}{2} + \frac{Y}{2}
\end{equation}
To obtain the charges $\{+,0\}$ in the components of $\phi$, one
has to assign $+1$ to $Y$, which turns out to be the {\it usual}
hypercharge of $\phi$.  In this way it becomes easier to
understand how $\phi$ transforms under $SU(2)$ but carries charges
of a larger symmetry.  The $SU(2) \times U(1)$ global symmetry of
the scalar field is implied\footnote{With the phrase $implied$ we
mean that the real scalar fields of $SO(4)$  can be cast in a
complex field of $U(2)$ . Note that this is not an embedding. We
think that in this way can the global charges of $SO(4)$ be taken
over by the global $SU(2) \times U(1)$ symmetry. The same applies
to the {\it currents} as well. Note that non-trivial currents
require complex conjugation, and make sense only within unitary
representations.} by $SO(4)$.

After that the scalar field is restated within a complex
representation of the $SU(2) \times U(1)$, we expect that, this
symmetry holds also locally, and describes gauge interactions with
locally conserved charges \footnote{Since the scalar fields are
local fields the $\gamma$ basis spanning $\phi$ should be
spacetime dependent as well, consequently the $SO(4)$ symmetry
should be a local symmetry {\it but not } a local gauge symmetry,
so that a local metric is defined.}. The partial derivatives in
eq. (\ref{scalarfield-c}) should be replaced with the gauge
covariant ones to assure local gauge invariance with respect to
the local gauge symmetry.

\section{Complexification}
Prior to any spontaneous symmetry breakdown and complexification,
where the latter will be clearly defined towards the end of this
section,  we assume that the scalar field $\phi$ in
eq.(\ref{vector}) transforms under $SO(4)$. Therefore
\begin{equation}\label{euclidean-S}
      \phi^T \phi = \frac{1}{4}
      \left(\phi^2_1+\phi^2_2+\phi^2_3+\phi^2_0\right)
\end{equation}
is an $SO(4)$ invariant of the scalar field. Prior to any
spontaneous symmetry breakdown and complexification the null
vector which is the {\it unstable} ground state, describes a {\it
false} vacuum and is trivially $SO(4)$ invariant. We {\it
additionally} assume that space-time, prior to any spontaneous
symmetry breakdown and complexification, transforms under $SO(4)$
as well.

From the other side we know that, in the spontaneously broken
phase of the local $SU(2) \times U(1)$ electroweak symmetry,
space-time transforms under $SO(3,1)$. Owing to our assumption
that initially both the scalar field and space-time are
transforming under $SO(4)$, we $expect$ analogously, that in the
spontaneously broken phase, the scalar field transforms under
$SO(3,1)$.i.e., it transforms like spacetime. Therefore in the
broken phase
\begin{equation}\label{minkowksi-S}
   \phi^T \phi = \frac{1}{4}
   \left(\phi^2_1+\phi^2_2+\phi^2_3-\phi^2_0\right)
\end{equation}
should be an $SO(3,1)$ invariant of the scalar field. The stable
ground state of the scalar field corresponds to the $true$ vacuum.
Consequently, the true vacuum should also satisfy the above
$SO(3,1)$ invariant relation.

We postulate that at some stage prior to spontaneous symmetry
breakdown, the scalar field must have undergone a phase
transition. Note that the transformation
\begin{equation}\label{phase_transition}
    \phi_0 \leftrightarrow i \, \phi_0
\end{equation}
switches us back and forth between eq.(\ref{euclidean-S}) and
(\ref{minkowksi-S}). From the other side, it is formally possible
to absorb the phase into the basis, in this way the transformation
switches us also back and forth between  the two sets of gamma
matrices given in Eq.(\ref{set1}) and (\ref{set2}). This can be
clarified in the following: The complex form of the scalar fields
in eq.~(\ref{doublet+0}) and (\ref{doublet-0}) are only maintained
if the component $\phi_0$ has  correctly $i$ as a prefactor. If we
let $ \phi_0 \rightarrow i \, \phi_0$ then the expansion in
eq.(\ref{thefour}) should be done over the basis
$\{\gamma_i,\gamma_0\}$ instead $\{\gamma_i,\tilde{\gamma}_0\}$.
This signals us that the initial $SO(4)$ and final $SO(3,1)$
symmetries of the scalar field are related over the phase of
$\phi_0$. The respective two sets $\{\gamma_i,\tilde{\gamma}_0\}$
and $\{\gamma_i,\gamma_0\}$ specify the initial and final metric
of spacetime. We will come back to this point later again.

We know that three components of the scalar field are associated
with the massless goldstone modes, which can be gauged away ( or
parameterized ) with an $SU(2)$ transformation. Since $SU(2)$ is
isomorphic to $SO(3)$, it is appropriate to assign the $
\phi_1,\phi_2,\phi_3$ fields to $SU(2)$.

The fourth field $\phi_0$ is associated with the Higgs field, we
know that the vacuum even after symmetry breakdown preserves the
local gauge symmetry in a hidden way, so the $U(1)$ piece of
$SU(2)\times U(1)$ should be related with the leftover field
$\phi_0$. Note that, in contrast to the usual assignment in the
Higgs mechanism implemented in the electroweak theory, the vacuum
expectation value will not be assigned here to $\phi_3$ but to
$\phi_0$, which is a major difference\footnote{The field $\phi_0$
is by definition initially real, but the $U(1)$ symmetry requires
complex representations. This is another hint to see why the
scalar field is apt to undergo complexification.}. This choice
covers interesting properties as will be later seen. Let us
consider the vacuum at a particular minimum resulting from eq.
(\ref{scalarfield-c}) such that
\begin{equation}\label{min}
\begin{split}
 & \phi_1=\phi_2=\phi_3=0 \\
 & \phi_0 = \frac{\sqrt{2}\mu}{\sqrt{\lambda}}= v \\
 & \phi_{\text{o}}=\left( \begin{array}{c}  0 \\
 i \, \frac{v}{\sqrt{2}}\\
\end{array} \right)
=\frac{v}{\sqrt{2}} \left( \begin{array}{c}  0 \\
 i \\
\end{array} \right)
\end{split}
\end{equation}
In the rest of this paper, the scalar fields
$\{\phi_1,\phi_2,\phi_3,\phi_0\}$ will be parameterized with
$\{\xi_1,\xi_2,\xi_3,\xi_0\}$ respectively. All degrees of freedom
should be linearly independent and the vacuum should be
consistently parameterizable. The above mentioned phase transition
described by $\phi_0 \rightarrow i\, \phi_0$ will turn out to be a
necessity for a consistent parametrization of the vacuum or more
generally of the scalar field. Two cases of parametrization are of
interest :

\par{\it Case I :} The scalar field is supposed to
transform under the local gauge symmetry, therefore the
parametrization of the scalar field around the vacuum state should
be done over the generators of the gauge group $\{\sigma_i,Y\}$
together with the expansion parameters of the scalar field
$\{\xi_1,\xi_2,\xi_3,\xi_0\}$ respectively, which closely amounts
to the so called unitary gauge\footnote{In principle this
parametrization is like that in the unitary gauge but differs in
the assignment of the expansion parameters as we discussed before.
See also eq. (\ref{c1})}.

\par{\it Case II :} The scalar field is supposed to transform
under the  $SO(4)$ symmetry, where the $\gamma$ basis has
coordinate dependence. Consistency requires that the scalar field
should also be parameterizable around the vacuum state through the
basis $\{\gamma_i,\tilde{\gamma}_0 \}$, with the expansion
parameters $\{\xi_1,\xi_2,\xi_3,\xi_0\}$ respectively. This latter
parametrization might suitably called the metric gauge.

Trough comparing the former and the latter parametrizations, we
explore the underlying condition that nature imposes on the scalar
field. The condition will turn out to become what we call {\it
Complexification}, and the local Euclidean symmetry will undergo a
change of signature.

\par{\it Parametrization - Case II :} Let us start with the
latter parametrization described in case II, by considering the
following ground state and the exponentiated transformation ;
\begin{equation}\label{Transformation}
 e^{- i \, \gamma \cdot \xi(x) / v - i\, \tilde{\gamma}_0 \cdot \xi_0/v}
\cdot \left( \begin{array}{c}
  0 \\
 -i  \frac{v}{\sqrt{2}}\\
  0 \\
  +i  \frac{v}{\sqrt{2}}\\
\end{array} \right)
\end{equation}
The exponential part if expanded up to first order reads
\begin{equation*}
    \left(%
\begin{array}{cccc}
  1  &  0  & -i\frac{\xi_3}{v}-\frac{\xi_0}{v}     & -i \frac{\xi_1-i \, \xi_2}{v}  \\
 0  & 1 & -i \frac{\xi_1 + i \, \xi_2}{v} & i\frac{\xi_3}{v}-\frac{\xi_0}{v}\\
 i\frac{\xi_3}{v} -\frac{\xi_0}{v} & i \frac{\xi_1 - i \, \xi_2}{v} & 1& 0\\
  i \frac{\xi_1 + i \, \xi_2}{v} & -i\frac{\xi_3}{v}-\frac{\xi_0}{v} & 0 & 1 \\
\end{array}%
\right)
\end{equation*}
the product of the vacuum state with the above expansion yields
\begin{equation*}
    \frac{1}{\sqrt{2}} \left(%
\begin{array}{c}
   \xi_1 - i \xi_2 \\
   -\xi_3 - i (\xi_0 + v)   \\
   \xi_1 - i \xi_2 \\
   -\xi_3 + i (\xi_0 + v)   \\\
\end{array}%
\right)
\end{equation*}
It is remarkable to see here that $\gamma_0$ behaves {\it
formally} like the hypercharge $Y$, and $\xi_0$ plays the role of
the Higgs field $H$, because it correctly appears beside $v$
within the parametrization. Let us consider the last term in the
exponential to investigate how the hypercharge acts:
\begin{equation}\label{hypercharge}
\begin{split}
 i \, \tilde{\gamma}_0 \cdot \xi_0  = i \, (-i \, \gamma_0 ) \cdot
 \xi_0  & =   \gamma_0  \cdot  \xi_0 \\
 & \equiv  -Y \cdot  \xi_0 \\
 & = i \, Y \cdot  ( i \, \xi_0 ) \\
 & = i \, Y \cdot  \tilde{\xi}_0  \\
\end{split}
\end{equation}
It is seen in the last line that $Y$ selects out $\tilde{\xi}_0
\stackrel {def}{=}i\xi_0 $ as the expansion parameter. Using the
same vacuum state, we verify this choice in the forthcoming
parametrization mentioned in case I.

\par{\it Parametrization - Case I :}
First we demonstrate how the parametrization works, then we
identify the $H$ field. Let us consider the following
transformation acting on the minimum
\begin{equation}\label{c1}
\phi(x) \approx \phi_0(x)= e^{- i \, \sigma \cdot \xi(x) / v - \,
i \, Y \cdot \xi_0/v} \ \ \left(
\begin{array}{c}
  0 \\
 i  \frac{v }{\sqrt{2}}\\
\end{array} \right)
\end{equation}
Expanding the exponential up to first order gives
\begin{equation*}
\begin{split}
&
\left(%
\begin{array}{cc}
   1-i\frac{\xi_3}{v}-i\, \frac{\xi_0}{v}  &  -i \frac{\xi_1 - i \, \xi_2}{v} \\
   -i \frac{\xi_1 + i \, \xi_2}{v} &  1+i\frac{\xi_3}{v}-i\,\frac{\xi_0}{v} \\
\end{array}%
\right) \left(
\begin{array}{c}
  0 \\
 i  \frac{v}{\sqrt{2}}\\
\end{array}\right)
\\ &
= \frac{1}{\sqrt{2}}\left(%
\begin{array}{c}
   \xi_1 - i \xi_2 \\
   -\xi_3 +\xi_0+ i v    \\
\end{array}%
\right)
\end{split}
\end{equation*}
It is seen that $\xi_0$ emerges at the wrong place and does not
add up to $v$, consequently doesn't operate like the $H$ field.
Since we demand that $\xi_0$ should operate\footnote{This
requirement might be understood in the context of {\it consistent
parametrization} of the scalar field around the vacuum.} like $H$,
we consider the possibility of  a phase transition $\xi_0
\rightarrow i\, \xi_0 = \tilde{\xi}_0$ which  was previously
postulated in eq. (\ref{phase_transition}). Note that this phase
transition can be compensated in $\tilde{\gamma}_0 \cdot \xi_0
\rightarrow -\, \gamma_0\cdot\tilde{\xi}_0 $, which preserves
invariance, and fulfills the previously stated condition in eq.
(\ref{hypercharge}). If we substitute $\tilde{\xi}_0$ back in the
last line above we obtain the correct form
\begin{equation*}
\frac{1}{\sqrt{2}}\left(%
\begin{array}{c}
   \xi_1 - i \xi_2 \\
   -\xi_3 + i (\xi_0 + v)    \\
\end{array}%
\right)
\end{equation*}
The phase transition is essential and leads the scalar field to
undergo a complexification, thereby spacetime changes signature
and becomes Minkowski. The four scalar fields should then
subsequently be reexpressed as
\begin{equation}
  \phi = \frac{1}{\sqrt{4}}\left( \begin{array}{c}
  \phi_1 \\
  \phi_2 \\
  \phi_3 \\
 i\phi_0 \\
  \end{array} \right)
\end{equation}
Consequently the scalar field satisfies the condition in eq.
(\ref{minkowksi-S}). Physically we will observe the expectation
value of $\phi_0$ and not its phase, the phase is swallowed by the
basis which becomes, $\{\gamma_0 , \gamma_1, \gamma_2, \gamma_3
\}$ and exhibits a Minkowski signature
\begin{equation*}
     2\, \eta_{\mu \nu} = -\{\gamma_\mu,\gamma_\nu\} \ , \ \ \ \
     \text{( $\mu,\nu$ = 0,1,2,3 )}
\end{equation*}
Since the scalar field is coordinate dependent, both
parameterizations should hold locally. Therefore the gamma basis
should have a coordinate dependence as well, and thereby defines a
local spin $2$ field through $\eta_{\mu \nu}$, which implies the
incorporation of general relativity. From the other side, in our
analysis, $SO(4)$ had naturally implied the  $U(2) \equiv SU(2)
\times U(1)$ unitary representation of the scalar field, giving
rise to local spin 1 gauge fields, which are the vector gauge
bosons mediating interactions. Finally the Spinorial
representation itself can naturally accommodate spinors, so are
the spin $1/2$ fields also incorporated.

\section{Lorentz Invariance}
In the unitary gauge, the photon has no couplings to the Higgs
field, nor does it have a mass term, provided over the vacuum
expectation value. As a result the photon does encounter no
resistance of the Higgs vacuum and the surviving $U(1)_{e}$
symmetry is Lorentz invariant.

In contrast, the $W$ and the $Z$ have couplings to $H$ and have
also mass terms. They are in contact with vacuum over tadpoles.
The Lorenz invariance of weak interactions reflect somehow the
following picture:

\textsl{If one were given the task to show that space-time is
Lorentz invariant {\it only} by using the massive $W$ and $Z$'s as
mediators of interactions, he or she would probably found out that
there is a dominant $SO(3)$ symmetry of space and an independent
measure of time, in the first place.}

Presumably in a world made up of massive mediators, which would
dramatically lead to a dominantly mechanical environment, that
means forces are short ranged and appear through contact,  it
wouldn't be possible to find out any trace of Lorentz Invariance.

The velocity of massive $W$ and $Z$'s are obviously no invariants
with respect to different inertial observers, whereas the photon
speed in vacuum is. This is told us by the coordinate
transformations that leave the Maxwell equations Invariant.
Another invariant property of the scalar field after becoming
massive, is its (rest) mass. Therefore we can work out  the vacuum
state $\phi_o$ by plugging it in eq. (\ref{inv}). We obtain :
\begin{equation}
\begin{split}
 -Tr\left[\, \phi \, \phi\, \right] = -Tr\left[ \left(\sum_i
 \frac{\phi_i \gamma_i}{4}
  - \frac{\gamma_0 (i\, \phi_0)}{4} \right)^2 \right]\\
  = \frac{1}{4}\left(\phi^2_1+\phi^2_2+\phi^2_3+\phi^2_0 \right)
 \stackrel{vac}{=}\frac{v^2}{4} = m^2_H c^4\\
\end{split}
\end{equation}
where  $m_H c^2$ is the invariant mass term of the Higgs field, at
the ground state where $\phi_1=\phi_2=\phi_3=0$ and $\phi_0=v$.
From $v= \frac{\sqrt{2}\mu}{\sqrt{\lambda}}$ and $m^2_H = \mu^2$ one
finds that $\lambda=\frac{1}{2}$. The potential function of the
scalar field prior to spontaneous symmetry breakdown and
complexification turns out to reduce in a natural form
\begin{equation}
 V(\phi) = -\frac{1}{2}m^2_H (\phi^\dagger \phi)
 + \frac{1}{8} (\phi^\dagger\phi)^2
\end{equation}
The Higgs mass $m_H$ is thereby determined as $m_H =  v/2 $
$\cong$ $ 123$ GeV. Actually the preceding relation should be
understood the other way around, conversely $m_H$ and $v$
determine the speed of light in the Higgs vacuum. This follows
from a simple consideration that we can normalize the vacuum state
$\phi_o$ in eq. (\ref{min}), into a unit state
\begin{equation}
 \left( \begin{array}{c}  0 \\
 i \\
\end{array} \right)
\end{equation}
by absorbing the factor $\sqrt{\frac{v}{\sqrt{2}}}$ into $\gamma_0
$ of the basis $\{ \gamma_1,\gamma_2,$ $\gamma_3,\gamma_0 \}$,
that spans $\{ \phi_1,\phi_2,\phi_3,i \,\phi_0 \}$. Furthermore
the invariant quantity $ Tr\left[\, \phi \, \phi\, \right]$
normalizes to $1/4$ ( in SI units to $c^2/4$ ), {\it if} we
further absorb $\sqrt{\frac{1}{\sqrt{2}m_H}}$ into $\gamma_0 $.
Thereby we redefine $\gamma_0$ for the sake of obtaining a unit
vacuum state such that
\begin{equation}
\begin{split}
    \gamma_0 \longrightarrow \left(\frac{v}{2 \,
    m_H}\right)^{\frac{1}{2}} \cdot \gamma_0
\end{split}
\end{equation}
The Minkowski signature reveals this  non uniformity (in SI units)
\begin{equation}
    \eta_{\mu \nu} =  \left(%
\begin{array}{cccc}
  1 & 0 & 0 & 0 \\
  0 & 1 & 0 & 0 \\
  0 & 0 & 1 & 0 \\
  0 & 0 & 0 & -c^2 \\
\end{array}%
\right)
\end{equation}
where the last entry is departing from $1$, and normalizes the
fourth spacetime component ; $x_0= i \, c \, t$. The factor
determines the speed of electromagnetic disturbances in the Higgs
vacuum :
\begin{equation}
 c  = \left(\frac{v}{2 \, m_H}\right)^{\frac{1}{2}}
 \ \ \ \ \ \  \text{: Speed of Light in Vac.}
\end{equation}
After the rescaling of $\gamma_0$, which follows from the
normalization of the vacuum state, we get
\begin{equation}\label{norm}
  -Tr\left[\, \phi \, \phi\, \right] =\frac{1}{4}\\
\end{equation}
This equality should always hold even when the Higgs field becomes
excited and departs from the ground state. In this respect the
parametrization allows us to utilize the $\xi_1,\xi_2,\xi_3$
fields, which can be chosen such that the invariant $-\phi^T \phi
=1/4$ is satisfied for any arbitrary value of $\xi_0\neq0$, which
is by definition the $H$ field. A suitable gauge would be ;
\begin{equation}
 \xi^2_0-\sum^3_{i=1} \xi_i^2  = 1 ; \ \ \ \ \xi_0= H ;
 \ \ \ \ \xi_0(0)=1
\end{equation}
The excited Goldstone modes $\xi_1,\xi_2,\xi_3$  can be gauged
away in the unitary gauge when necessary.

\par{\it Complexity of the Vacuum :} The time variable $t$ is real
and thus measurable, but formally can be made to enter the Lorentz
transformations with a complex sign, so that the rotation is over
a complex angle in a complex plane. A similar situation arose in
the vacuum. The ground state in eq. (\ref{min}) contains a complex
sign. However the vacuum expectation value $v$ is itself real and
is gained by the component $\phi_0$. The ground state enters the
transformation with a complex prefactor just like $t$.

\section{Conclusion}
We shortly highlight here the underlying steps that lead to the
complexification of the vacuum :
\begin{enumerate}
    \item[(a)] The real valued scalar field, can be cast in a
    complex representation, so that the global $SO(4)$ invariant
    Lagrangian naturally implies a global $U(2)$ invariant Lagrangian.
    \item[(b)] The global charges remnant of $SO(4)$ are taken over
    by the unitary representation $U(1) \times SU(2)$.
    This symmetry turns out to hold as a local gauge symmetry. The
    gamma basis should also be spacetime dependent.
    \item[(c)] A consistent parametrization
    through the unitary and metric gauges
    assigns the real fields $\phi_1,\phi_2,\phi_3$ to $SU(2)$
    and $\phi_0$ to $U(1)$ and also requires the complexification
    of the scalar field; $\phi_0 \rightarrow i \phi_0$ which
    amounts to a change in signature of spacetime;
    $i \, \gamma_0 \rightarrow  \gamma_0$
    \item[(d)] The scalar field develops an invariant mass term
    through the spontaneous breakdown of the unstable ground state
    to the stable ground state. The spontaneous breakdown is
    likely to be induced by the complexification;
    Since the product $-\mu^2 (\phi^T \phi) $, contributes for
    $\phi_0 \rightarrow i \phi_0$ a mass term with the correct sign.
    \item[(e)] Normalization of the true vacuum state
    places a constant factor $c=\sqrt{\frac{v}{2 \,m_H}}$ in
    front of $\gamma_0$. The rescaled  gamma basis inherently defines
    the speed of light in the Higgs vacuum, over the $SO(3,1)$
    invariance of the Maxwell equations.
\end{enumerate}

\end{document}